\documentstyle[aps,twocolumn,epsf]{revtex}  
\begin{document}
\twocolumn[\hsize\textwidth\columnwidth\hsize\csname 
@twocolumnfalse\endcsname                            
\title{Liquid-Gas phase transition in Bose-Einstein Condensates with
time evolution }
\author{A. Gammal,$^{1}$ T. Frederico,$^{2}$ Lauro Tomio,$^{1}$ 
and Ph. Chomaz$^{3}$}
\address{$^{1}$ Instituto de F\'{\i}sica Te\'{o}rica, 
Universidade Estadual Paulista, 01405-900,  S\~{a}o Paulo, Brazil \\
$^{2}$Departamento de F\'{\i }sica, Instituto Tecnol\'{o}gico da
Aeron\'{a}utica, CTA,
12228-900, S\~{a}o Jos\'{e} dos Campos, Brazil \\
$^{3}$ GANIL, B.P. 5027, F-14021 Caen Cedex, France }
\maketitle
\begin{abstract}
We study the effects of a repulsive three-body interaction on a system of
trapped ultra-cold atoms in Bose-Einstein condensed state. The stationary 
solutions of the corresponding $s-$wave non-linear Schr\"{o}dinger equation 
suggest a scenario of first-order liquid-gas phase transition in the
condensed state up to a critical strength of the effective three-body
force. The time evolution of the condensate with feeding process and 
three-body recombination losses has a new characteristic pattern. 
Also, the decay time of the dense (liquid) phase is longer than expected
due to strong oscillations of the mean-square-radius.
\newline{PACS 03.75.Fi, 36.40.Ei, 05.30.Jp, 34.10.+x}
\end{abstract}
\vskip 0.5cm ]                              

The experimental evidences of Bose-Einstein condensation (BEC) in
magnetically trapped weakly interacting atoms~\cite{and96,mew96,brad97}
brought a considerable support to the theoretical research on bosonic
condensation. The nature of the effective atom-atom interaction determines
the stability of the condensed state: the two-body pseudopotential is
repulsive for a positive $s-$wave atom-atom scattering length and it is
attractive for a negative scattering length~\cite{huang}. The ultra-cold
trapped atoms with repulsive two-body interaction undergoes a Bose-Einstein
phase-transition to a stable condensed state, in a number of cases found
experimentally, as for $^{87}$Rb~\cite{and96}, for $^{23}$Na~\cite{mew96}
and $^{7}$Li~\cite{brad97}. \ However, a condensed state of atoms with
negative $s-$wave atom-atom scattering length would be unstable for a large
number of atoms \cite{rup95,baym96}.
It was indeed observed in the $^{7}$Li gas~\cite{brad97}, for which the 
$s-$wave scattering length is $a=(-14.5\pm 0.4)$ \AA , that the number of
allowed atoms in the condensed state was limited to a maximum value between
650 and 1300, which is consistent with the mean-field prediction~\cite{rup95}. 


From a theoretical approach, the addition of a repulsive three-body
interaction can extend considerably the region of stability for a
condensate even for a very weak three-body force~\cite{ADV-GFT}. 
As one can observe from Refs.~\cite{MV}, both signs for the 
three-body interaction are, in principle, allowed. However, 
in the present study we only consider the  case
of a repulsive three-body elastic interaction together with
an attractive two body interaction. We will show that,
due to the repulsive three-body force, new physical aspects
appears in the time evolution of the condensate.
In respect to the static situation, 
it was suggested that, for a large number of bosons the three-body repulsion
can overcome the two-body attraction, and a stable condensate will appear in
the trap~\cite{josse}. \ Singh and Rokhsar\cite{SR} have also
observed that above a critical value 
the only local minimum is a dense gas state, where the
neglect of three-body collisions fails. 

In this work, using the mean-field approximation, we 
develop the scenario of collapse, which includes two aspects of
three-body interaction, that is recombination and replusive mean-field
interaction. We begin by investigating the
competition between the leading term of an attractive two-body interaction,
which is originated from a negative two-atom $s-$wave scattering length, and
a repulsive three-body interaction, which can happen in the 
Efimov limit~\cite{efimov}, when $|a|\rightarrow \infty $.
(The physics of three-atoms in the Efimov limit is discussed in
Refs.~\cite{3atom}). 
We first consider the stationary solutions of the corresponding extension
of the Ginzburg-Pitaevskii-Gross (GPG)~\cite{gin} nonlinear Schr\"odinger
equation (NLSE), for fixed  number of particles, without dissipative terms, 
extending an analysis previously reported in Refs.~\cite{ADV-GFT}.
The liquid-gas phase transition in the condensate, suggested in 
\cite{ADV-GFT}, was confirmed by a more detailed analysis in the present 
stationary calculations. \ Then, the time evolution of the feeding process of
the condensate  by an external source is obtained by solving 
the time-dependent NLSE with repulsive three-body interaction 
(given by $g_3>0$) and dissipation  due to three-body recombination
processes. The dramatic collapse and the consequent atom loss that
happens at the critical number of atoms (when $g_3=0$)~\cite{SSH} is
softened by the addition of the three-body repulsive force. The decay
time of the liquid-phase is also unexpectedly long, when compared with
the decay time that occurs for $g_3=0$,  
which gives a clue about the possible observation of three-body 
interaction effects. \ Our results pointed out that the 
mean\--square\--radius is an important observable to be analyzed
experimentally to study the dynamics of the growth and collapse of the 
condensate~\cite{SSH}. 
\ In the present study, in order to enphasize
the real part of the three-body interaction, we choose 
$g_3$ significantly larger than the magnitude of the dissipative
term; although, in general, they are expected to be of the same order.

The NLSE, which describes the condensed wave-function in the mean-field
approximation, after considering the two-body attractive and 
three-body repulsive effective interaction, is variationally obtained from
the corresponding effective Lagrangian (see Gammal {\it et al.} in 
\cite{ADV-GFT}). By considering a stationary solution, 
$\Psi (\vec{r},t)=e^{-i\mu t/\hbar }$ $\psi (\vec{r})$
where $\mu $ is the chemical potential and $\psi (\vec{r})$ is normalized 
to the number of atoms $N$. By rescaling the NLSE for the $s-$wave
solution, we obtain 
\begin{equation}
\left[ -\frac{d^{2}}{dx^{2}}+\frac{x^2}{4}-\frac{|\phi
(x)|^{2}}{x^{2}}
+g_{3}\frac{|\phi (x)|^{4}}{x^{4}}\right] \phi(x)=\beta \phi (x) 
\label{schd}
\end{equation}
for $a<0$, where $x\equiv \sqrt{{2m\omega }/{\hbar }}\ r$ and 
$\phi(x)\equiv
\sqrt{8\pi |a|}\;r\psi (\vec{r})$. 
The dimensionless parameters, related to
the chemical potential and the three-body strength are, respectively,
given by $\beta \equiv \mu /\hbar \omega $ and $g_{3}\equiv \lambda _{3}\hbar
\omega m^{2}/(4\pi \hbar ^{2}a)^{2}$. The normalization for $\phi(x)$
reads $\int_{0}^{\infty }dx|\phi (x)|^{2}\ =n$ where the reduced number
$n$ is related to $N$ by $n\equiv 2N|a|\sqrt{2m\omega/\hbar }.$
The boundary conditions~\cite{rup95} in Eq.(\ref{schd}) are given by
$\phi(0)=0$ and $\phi (x)$ $\to C\exp (-x^{2}/4+[\beta -\frac{1}{2}]\ln
(x))$ when $x\to \infty .$~\cite{GFT2} 

\begin{figure}
\setlength{\epsfxsize}{1.0\hsize} \centerline{\epsfbox{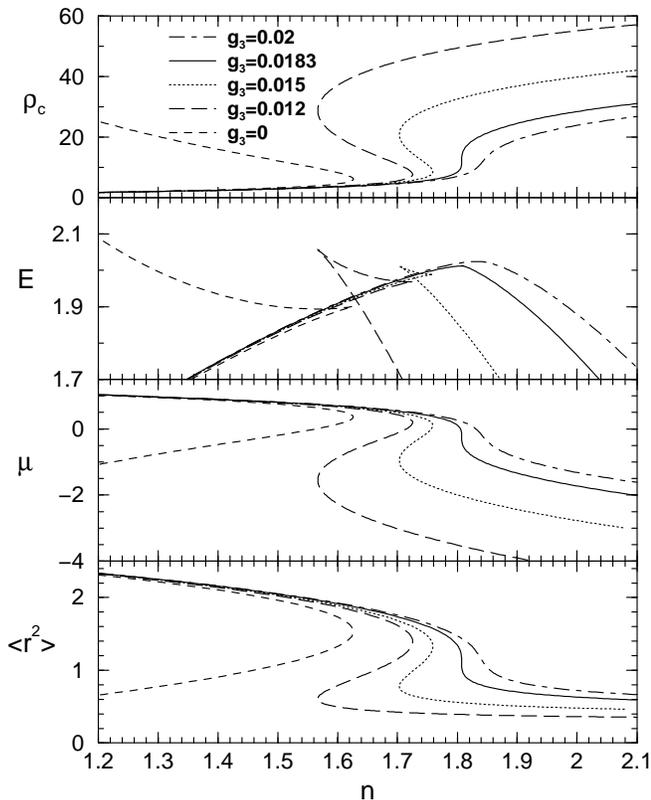}}
\caption{ Central density $\rho _{c}$, total energy $E$, chemical
potential $\mu $, and average square radius $\left\langle r^{2}\right\rangle 
$, as functions of the reduced number of atoms $n.$ 
The three-body strength $g_3$ are given in the upper frame. 
The corresponding units are: $(m\omega /\hbar )/(4\pi |a|)$ for $\rho
_{c}$, $ (N\hbar \omega )/n$ for $E$, $\hbar \omega $ for $\mu $, and
$\hbar/(2m\omega )$ for $\left\langle r^{2}\right\rangle $.}
\end{figure}
In Fig. 1, considering several values of $g_{3}$ (0, 0.012, 0.015, 0.0183
and 0.02), using exact numerical calculations, we present the evolution of some
relevant physical quantities $E,$ $\mu ,$ $\rho _{c}$ and $\ \left\langle
r^{2}\right\rangle $ as functions of the reduced number of atoms $n$.
\ For $g_{3}=0$, our calculation reproduces the result presented in 
Ref.~\cite{rup95,hs}, with the maximum number of atoms limited by 
$n_{max}\approx 1.62$ \ ($n$ is equal to $|C_{nl}^{3D}|$ of 
Ref.~\cite{rup95}). 

As shown in the figure, for $0<\ g_{3}\ <\ 0.0183$, the density $\rho _{c},$
the chemical potential $\mu $ and the root-mean-squared radius $\left\langle
r^{2}\right\rangle $ present back bendings typical of a first order phase
transition. For each $g_{3},$ the transition point given by the crossing
point in the $E$ versus $n$ corresponds to a Maxwell construction in the
diagram of $\mu $ versus $n$. At this point an equilibrated
condensate should undergo a phase transition from the branch extending to
small $n$ to the branch extending to large $n.$ The system should never
explore the back bending part of the diagram because as we have seen in
Fig. 1 it is an unstable extremum of the energy. From this figure it is
clear that the first branch is associated with large radii, small densities
and positive chemical potentials while the second branch presents a more
compact configuration with a smaller radius a larger density and a negative
chemical potential. This justify the term gas for the first one and liquid
for the second one. However we want to stress that both solutions are
quantum fluids. With $g_{3}=0.012$ the gas phase happens for $n<1.64$ and
the liquid phase for $n>1.64$. For $g_{3}\ >\ 0.0183$ all the presented
curves are well behaved and a single fluid phase is observed.
At $ g_{3}\approx 0.0183$ and $n\approx 1.8$, the stable, metastable and
unstable solutions come to be the same. This corresponds to a critical
point associated with a second order phase transition. At this point the
derivatives of $\mu ,$ $\rho _{c}$ and $\ \left\langle r^{2}\right\rangle $
as a function of $n$ all diverge.
We also checked that calculations with the variational expression of
$\langle r^2\rangle$, $\rho_c$ and $\mu$ are in good agreement with the
ones depicted in Fig. 1.

In the lower frame of Fig. 2, we show the phase boundary separating 
the two phases in the plane defined by $n$ and $g_{3}$ and the critical point 
at $n\approx 1.8$ and $g_{3}\approx 0.0183$. In the upper frame, we show the
boundary of the forbidden region in the central density versus  
$g_{3}$ diagram. 

\begin{figure}
\setlength{\epsfxsize}{1.0\hsize} \centerline{\epsfbox{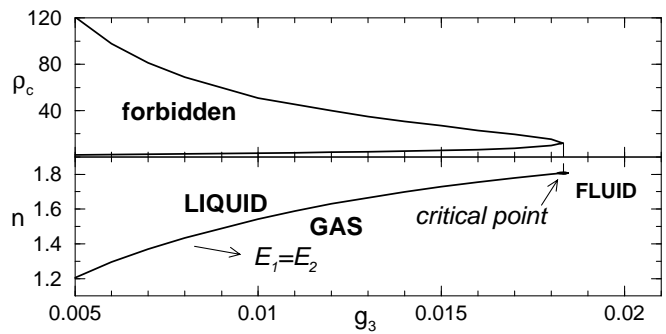}}
\caption{Phase diagram of the Bose condensate. Central density
$\rho_c $ \ in units of $(m\omega/\hbar)/(4\pi |a|).$}
\end{figure}

The main physical characteristic of the repulsive three-body force is to
prevent the collapse of the condensate for the particle number above the
critical number found with only two-body attractive interaction. The three-body
repulsive potential tends to overcome the attraction of the two-body potential
at short distances, as described by Eq. (3), as the repulsive 
interaction grows inversely to $x^4$, while the two-body potential 
is proportional to $x^{-2}$. Thus, the implosive force that shrinks 
the condensate at the critical number  is compensated by
the repulsive three-body force. The time evolution of the 
growth and collapse of the condensate with attractive
interactions~\cite{SSH} should be qualitatively
modified by the presence of the repulsive three-body force. 
The three-body recombination effect~\cite{KMS}, which ``burns" partially 
the condensed state should be taken into account
to describe quantitatively the dynamics of the condensate. 
In the case of only two-body attractive interaction, 
as observed by Kagan et al.~\cite{KMS}, 
by considering the feeding of the condensate from the nonequilibrium 
thermal cloud, the time evolution is dominated by a sequence of 
growth and collapse of the trapped condensate. 
The collapse occurs when the number of atoms in the 
condensate exceeds the critical number $N_c$; 
and it is followed by an expansion after the atoms in the high density 
region of the wave-function are lost due to three-body recombination 
processes and consequently the average attractive potential from the 
two-body force is weakened.
It is also noticed in Ref.~\cite{KMS} that the condensate time evolution is 
dominated by an oscillatory mode of frequency $\sim\omega$; and,
as time grows and $N$ reaches a value $> N_c$, 
a huge compression takes place to implode the condensate. 
The repulsion given by the three-body force will dynamically affect
the compression of the condensate weakening the implosive force and
allowing more atoms to survive at high densities.

In order to quantitatively study the above features with repulsive three-body 
interaction, we consider the time-dependent non-linear Schr\"odinger
equation corresponding to Eq.~(\ref{schd}), including three-body recombination 
effects (with an intensity parameter $2\xi$) and an imaginary linear term 
corresponding to the feeding of the condensate (with intensity parameter 
$\gamma$):
\begin{eqnarray}
i\frac{\partial\Phi}{d\tau} 
&=&  
\left[ -\frac{d^{2}}{dx^{2}}+\frac{x^2}{4}-
\frac{|\Phi|^{2}} {x^{2}} +
(g_{3}-2i\xi) \frac{|\Phi|^{4}}{x^{4}} +
\frac{i\gamma}{2}
\right] \Phi 
\label{tschd} ,
\end{eqnarray}
where $\Phi\equiv \Phi(x,\tau)$ and $\tau\equiv\omega t$. 
For the parameters $\xi$ and $\gamma$ we
are using the same notation as given in Ref.~\cite{KMS}.

In Fig. 3, we show the time evolution of the number of condensed atoms,
starting with $N/N_c=0.75$, found by the numerical solution of 
Eq.~(\ref{tschd}) with $\xi=$0.001 and $\gamma=$0.1, with and without
repulsive three-body potential. We compare  the results of a three-body
potential with $g_3=$ 0.016 to the case considered in Ref.~\cite{KMS},
with $g_3=$0. In both, $N_c$ is the critical number for $g_3=$0. 
The first striking feature with repulsive three-body force 
is the smoothness of the compression mode in comparison with the 
results of $g_3=0$. This is a result of the explosive force from the
repulsion, which oppose to the sudden density increase and damps the loss
of atoms due to three-body recombination effects. Even for $g_3$ lower 
than 0.016, and much closer to $g_3=0$, the collapses can no longer
``burn'' the same number of atoms as in the case of $g_3=0$.
By extending our calculation presented in Fig. 3 for all cases
with $g_3>0.01$ and for times beyond $\omega t= 50$,
we have checked that the number of atoms will increase without limit 
while the condensate is oscillating with frequency about $2\omega$.
In particular, the present approach indicates that the experimental 
recent observation of the maximum number of $^7$Li atoms is compatible 
with $g_3$ much smaller than 0.01. 
The mean square radius for $g_3=0$, after each strong collapse 
(when $N>N_c$) begins to oscillate at an increased average radius. 
The collapse ``burns" the atoms in the states with higher densities and
explain the sudden increase of the square radius after each compression,
remaining the atoms in dilute states. The inclusion of the repulsive 
three-body force, still maintains the oscillatory mode, but the compression
is not as dramatic as in the former case, and consequently atoms in  
higher density states are not so efficiently burned. The increase of the 
mean square radius (averaged with time) is smaller than the one
found with only attractive two-body force.
This is a remarkable feature of the stabilizing effect of the
repulsive three-body force allowing the presence of states with higher
densities, as we found in the stationary study. 

\begin{figure}
\setlength{\epsfxsize}{1.0\hsize} \centerline{\epsfbox{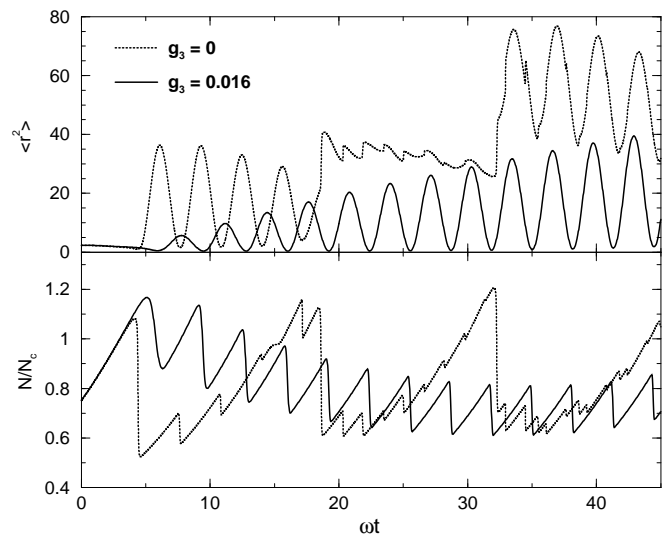}}
\caption{Number of condensed atoms and the
corresponding mean square radius $\langle r^2 \rangle$ (in units of  
$\hbar/(2m\omega)$) as a function of time, for $\xi=$0.001, $\gamma=$0.1.
$N_c$ is the maximum number of atoms for $g_3=0$.}
\end{figure}

Finally, we have to consider that, in the situation when the 3-body 
repulsion dominates over the 2-body attraction, the condensate
can be in a denser phase where it is expected to be strongly unstable 
due to recombination losses. The decay time of the condensate in a 
denser phase is expected to be much smaller than the decay time
of the condensate in the less dense phase. However, we should observe 
that the dynamics of the condensate is modulated by an oscillatory mode
with a frequency of the order of $2\omega$, which was already 
identified by \cite{KMS} to be $\sim\omega$ even when $g_3 =0$. 
In case of $g_3 > 0$, such oscillatory mode dominates the time 
evolution of the condensate. As the oscillations allow changes 
in the density, the condensate does not ``burn" as fast as expected.

In order to study the condensate decay, we consider the original
NLSE with the dissipative term and allow different possibilities for the
three-body interaction $g_3$. We use  $\xi = 0.001$ (the same value used
in Ref.~\cite{KMS}). In Fig.~4 we show the result of this study for
$g_3=0$ and $g_3=0.016$. The initial number of atoms $N$ can be obtained from
$n$, given in the figure.  For $g_3=0$, we took $n=1.625$, which is close
to the critical limit. For $g_3=0.016$, we consider three cases: two of
them starting with the same number of atoms, $n=1.756$, but in different 
phases (the corresponding chemical potentials are $\beta=-1.2$, in a denser 
phase, and $\beta=0.3$); and another in an even denser phase, with 
$n=1.965$ and $\beta=-2.3$  (see also Fig.~1).
Based on the results obtained in these four different cases, we can
estimate that the mean-life for the condensate, which is initially 
in a denser phase, is not as small as expected when comparing with $g_3=0$. 
We observe in this case the relevant role of the oscillatory mode,  
related to the frequency of the trap potential, which  
dominates the dynamics of the condensate when $g_3 > 0$.

\begin{figure}
\setlength{\epsfxsize}{1.0\hsize} \centerline{\epsfbox{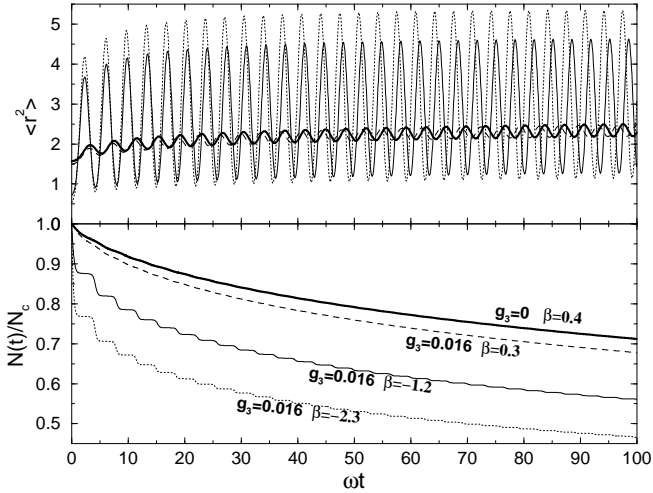}}
\caption{Condensate decay. Number of condensed atoms
and mean square radius as function of $\omega t$, for $\gamma=0$ and 
$\xi=0.001$. The chemical potential $\beta$ of the initial state 
and the strengths $g_3$ are given in the plot.}
\end{figure}

To summarize, our present results can be relevant to determine a possible
clear signature of the presence of a repulsive three-body interaction in
Bose condensed states. It points out to a new type of phase transition
between two Bose fluids. Because of the condensation of the atoms in a
single wave-function this transition may present very peculiar
fluctuations and correlation properties. As a consequence, it may fall
into a different universality class than the standard liquid-gas phase
transition, which are strongly affected by many-body correlations. 
The characterization of the two-phases through their energies, chemical
potentials, central densities and radius were also given for several
values of the three-body parameter $g_{3}$. 
We develop a scenario of collapse which includes both
three-body recombination and three-body repulsive interaction.
From  the time-dependent analysis, we 
show that the decay time of the condensate which begins in a denser 
phase is long enough to allow observation. However, the observed strongly
oscillating states are quite different from the analysed stationary 
states. In accordance  with the observed strong 
oscillations of the mean-squared-radius, the condensate density also 
strongly oscillates and the observed states cannot be characterized 
as ``dense" or ``dilute", justifying the long decay time.
Nevertheless, through the amplitude of the oscillations one can 
distinguis if the system starts in a denser phase.
\vskip 0.2cm


AG, TF and LT thank Profs. G.V. Shlyapnikov and A.E. Muryshev for
providing details of their numerical calculations; Prof. R.G. Hulet for 
details on experimental results and Profs. N. Akhmediev and M.P. Das 
for relevant correspondence related to Ref.~\cite{ADV-GFT}.
This work was partially supported by Funda\c c\~ao de Amparo \`a Pesquisa
do Estado de S\~ao Paulo and Conselho Nacional de Desenvolvimento
Cient\'\i fico e Tecnol\'ogico.

\end{document}